\documentclass[preprint,floats,floatfix,superscriptaddress,prl]{revtex4}

\usepackage{graphicx}
\usepackage{amsmath}
\usepackage{amsfonts}
\usepackage{bbm}
\usepackage[utf8]{inputenc}

\begin{document}

\title{Singular Values, Nematic Disclinations, and Emergent Biaxiality}

\author{Simon \v{C}opar}

\affiliation{Faculty of Mathematics and Physics, University of
Ljubljana, Jadranska 19, 1000 Ljubljana, Slovenia}
\affiliation{Department of Physics and Astronomy, University of Pennsylvania, 209 South 33rd Street, Philadelphia, Pennsylvania 19104, USA}

\author{Mark R. Dennis}

\affiliation{H. H. Wills Physics Laboratory, University of Bristol, Bristol BS8 1TL, United Kingdom}

\author{Randall D. Kamien}

\affiliation{Department of Physics and Astronomy, University of Pennsylvania, 209 South 33rd Street, Philadelphia, Pennsylvania 19104, USA}

\author{Slobodan \v{Z}umer}

\affiliation{Faculty of Mathematics and Physics, University of
  Ljubljana, Jadranska 19, 1000 Ljubljana, Slovenia}

\affiliation{J. Stefan Institute, Jamova 39, 1000 Ljubljana,
  Slovenia}

\date{\today}

\begin{abstract}
Both uniaxial and biaxial nematic liquid crystals are defined by
orientational ordering of their building blocks.  While uniaxial
nematics only orient the long molecular axis, biaxial order implies
local order along three axes. As the natural degree of biaxiality and
the associated frame, that can be extracted from the tensorial
description of the nematic order, vanishes in the uniaxial phase, we
extend the nematic director to a full biaxial frame by making use of a
singular value decomposition of the gradient of the director field
instead. New defects and degrees of freedom are unveiled and the
similarities and differences between the uniaxial and biaxial phase
are analyzed by applying the algebraic rules of the quaternion group
to the uniaxial phase.
\end{abstract}

\newcommand\ii{\mathbbm{i}}
\newcommand\jj{\mathbbm{j}}
\newcommand\kk{\mathbbm{k}}
\newcommand\bone{\mathbbm{1}}
\renewcommand\vec[1]{\mathbf{#1}}

\maketitle
 
Nematic liquid crystals are revered for their geometrically complex
and visually compelling defect structures, stabilized by topological
constraints \cite{degennes_book}. The elementary rules of homotopy
that govern these defects
\cite{kleman_defden,klem_mich_78,mermin,volovik_mineev,kurik_lav,trebin_rev,randy_rmp}
imply ambiguities in defect classification when disclination lines are
involved due to the action of the first homotopy group on itself and
on the second homotopy group. Nematic defects have been probed in both
nematic and cholesteric colloids and emulsions
\cite{gu_abbott,lubensky_pettey_stark,lavrentovich_worlds,musevic_science},
highly confined geometries \cite{serra_porous,
  serra_nmat_mem,fukuda_blue_prl}, and optically manipulated liquid
crystals \cite{musevic_ghosts,smalyukh_trapping}. So robust are they,
that they can be manipulated to reproducibly form linked or knotted
disclination lines
\cite{araki_collagg,ravnik_dimers_wires,uros_science,jampani_knots} as
well as other topologically interesting objects
\cite{uros_vortex,smalyukh_toron}. Recall that uniaxial nematics can
be interpreted as a highly symmetric special case of biaxial nematics
\cite{mermin,kurik_lav}, suggesting an opportunity to study nematic
defects with tools that are not available in the standard homotopy
theory.  In this article, we explore the similarity between distortion
patterns found in uniaxial nematic fields and defects in biaxial
phases by introducing a new biaxial frame {\sl derived entirely from
  deformations of the uniaxial director field $\vec{n}$}.  Like the
Frenet-Serret frame of a curve or the principal axes frame of a
surface \cite{kamien_primer}, our new frame has a well-defined
(differential) geometric meaning and allows us to provide a
topological characterization to the director geometry.  The
``quasi-defects'' in this new frame allow us to apply the
well-developed theory of biaxial nematics and to include the
non-topological ``escaped defect'' \cite{Meyer} in our classification,
embellishing, for instance, our understanding of the double-twist tube
construction \cite{bluephasedt} of the blue phases. A motivation for
this investigation is the study of similar quasi-defect structures in
optics, where topological filaments in the derivative of a complex
scalar field determine the topology of optical vortices
\cite{dennis_twirl, berry_dennis}.  We demonstrate our technique on
numerical models of blue phases and discuss the implications of newly
extracted information.

Uniaxial nematics consist of elongated non-polar molecules that tend
to align in a particular direction in space, taking directions in the
manifold $\mathbb{R}P^2$. The \emph{director} is specified by a unit
vector $\vec{n}$ up to sign. As a result, uniaxial nematics
accommodate both line defects (disclinations) and point defects
\cite{mermin,randy_rmp} since $\pi_1(\mathbb{R}P^2)=\mathbb{Z}_2$ and
$\pi_2(\mathbb{R}P^2)=\mathbb{Z}$. While point defects can be
oriented, a line defect, a disclination line, winds the director by
$\pi$ leading to a sign inconsistency.  Recall, however, that
non-defect states exist as disclinations with a non-topological
integer winding number. The famous escape into the third dimension
renders these smooth \cite{Meyer} in $\vec{n}$. As a line field alone,
the nematic director does not have an intrinsic biaxial nature. In
order to define a biaxial structure, we turn to the tensor of
gradients, $\partial_i n_j$, which provides the additional structure
necessary to define an entire frame. Multiplying from the left by a
unit vector extracts a directional derivative of $\vec{n}$; there are
special orthogonal directions $[\vec{w}^1,\vec{w}^2,\vec{w}^3]$, in
which the magnitudes $||w^\alpha_i \partial_i n_j||^2$ of the
derivatives are extreme, which can also be seen as an eigenvalue
problem. The derivatives in these directions take the form of
$w_i^\alpha \partial_i n_j=\sigma_\alpha n^\alpha_j$ for each
$\alpha\in \{1,2,3\}$, where $[\vec{n}^1,\vec{n}^2,\vec{n}^3]$ define
a {\sl different} orthonormal frame in the domain of the matrix in
such a way, that the singular values $\sigma_\alpha$ are positive
semi-definite and satisfy $\sigma_1\geq\sigma_2\geq\sigma_3\geq0$.
Because the director $\vec{n}$ is a unit vector, it is in the kernel
of the derivative tensor, $(\partial_in_j)n_j=0$, which implies
$\sigma_3=0$ and $\vec{n}^3\equiv \vec{n}$. This makes the axes
$\vec{n}^1$ and $\vec{n}^2$ orthogonal to $\vec{n}$ and decorate the
uniaxial phase with a biaxial order! We thus decompose the gradient
tensor as:
\begin{equation}
  \partial_in_j=\sigma_1 w^1_i n^1_j +\sigma_2 w^2_i n^2_j  \label{eq:svd}
\end{equation}
By construction, the derivative of the director is largest in the
direction $\vec{w}^1$, changing towards $\vec{n}^1$ with the rate of
$\sigma_1$. The derivative in the direction $\vec{w}^2$ points towards
$\vec{n}^2$ with a lower rate of $\sigma_2$. The remaining vector
$\vec{w}^3\equiv\vec{w}$ marks the direction in which the director is
constant. This is known as the singular value decomposition (SVD) of
the matrix $\partial_in_j$ and proves to be just the thing we need.

It is amusing to note that the derivative tensor is intimately
connected to the topological hedgehog charge through the Gauss
integral: \cite{randy_rmp,klem_lavr},
\begin{eqnarray}
  q&=&\frac{1}{4\pi}\iint
  \frac{1}{2}\epsilon_{ijk}\epsilon_{lmp}n_l\partial_j n_m \partial_k
  n_p {\,\rm d} S_i\nonumber\\
  &=&\frac{1}{4\pi}\iint\sigma_1\sigma_2\vec{w}\cdot{\,\rm d}\vec{S}.
  \label{eq:qdef}
\end{eqnarray}
Thus the streamlines of the vector field
$\tilde{\vec{w}}\equiv\sigma_1\sigma_2\vec{w}$ trace the preimage of a
particular director orientation as a three-dimensional generalization
of the schlieren texture \cite{chen} and the topological charge is
simply the flux of these streamlines through the enclosing
surface. The streamlines can only terminate on the singular nematic
defects or where the $\tilde{\vec{w}}$ field disappears, {\sl i.e.},
when $\sigma_2=0$. The latter are saddles in the derivative field,
where the director is constant in two directions.

The SVD frame decorates the director field with two new vectors that
encode the transverse degrees of freedom and can be interpreted as a
coordinate in ${\rm SO}(3)/D_2$ to parameterize biaxial order. In
regions where $\partial_i n_j$ does not vanish, the frame is
continuous everywhere except on a one-dimensional set of points, which
we collectively call line defects or disclinations in the SVD triad,
as they are analogous to disclinations in biaxial nematics. There are
two types of defects in this frame. On \emph{native disclinations} --
line and point defects in $\vec{n}$ (for simplicity, we can treat the
point defects as small disclination loops \cite{randy_rmp}), the
derivative tensor diverges and the frames and singular values are
ill-defined. Away from the defects in $\vec{n}$, the SVD decomposition
can still give a frame with degenerate axes wherever two singular
values coincide; we call these locations
\emph{quasi-disclinations}. Both the $\vec{n}$-frame and the
$\vec{w}$-frame share the defects, which are characterized by
singularities and degeneracies in the singular values. Each
disclination line is characterized by a \emph{pair} of degenerate axes
and by the amount of rotation of these axes around the remaining
nondegenerate axis -- the winding number of the disclination. We will
draw an analogy with the biaxial phase and its first homotopy group --
the quaternion group written in terms of unit quaternions,
$\{\bone,-\bone,\pm \ii,\pm\jj,\pm \kk\}$
\cite{wrightmermin,randy_rmp}. For the disclinations with half-integer
winding number, we shall assign a unit quaternion $\ii$ to a defect in
axes $\{\vec{n}^2,\vec{n}\}$, $\jj$ to a defect in
$\{\vec{n},\vec{n}^1\}$ and $\kk$ to a defect in
$\{\vec{n}^1,\vec{n}^2\}$. All disclinations with odd integer winding
number belong to the class $-\bone$, regardless of which pair of axes
they involve. Finally, the disclinations with even integer winding
number are all trivial, in the $\bone$ class.

\begin{figure*}
  \centering
  \includegraphics[width=\textwidth]{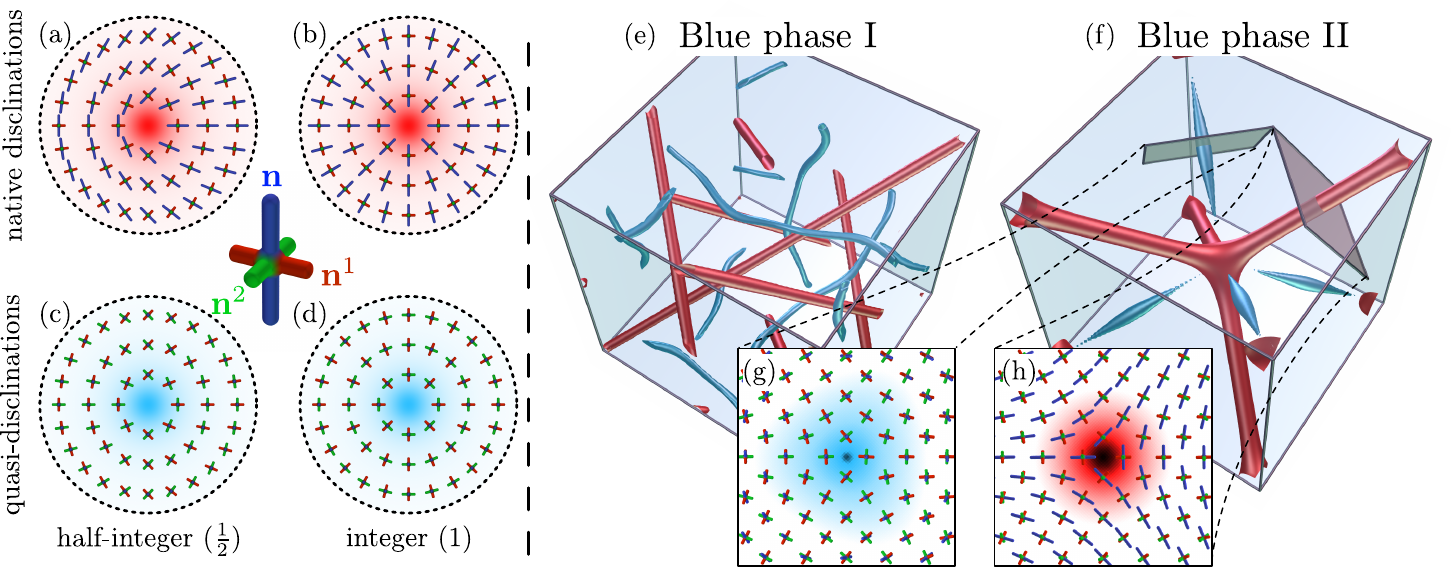}
  \caption{
    \label{fig:fig_blues}
    (color)
    Sketched profiles for both the native disclinations (a,b) and the
    hidden quasi-disclinations (c,d). The native nematic
    disclinations involve half-integer (a) or integer (b) rotation of
    the director $\vec{n}$ and the perpendicular axis $\vec{n}^1$. The
    quasi-disclinations are similar, except the auxiliary axes
    $\{\vec{n}^1,\vec{n}^2\}$ are degenerate instead of
    $\{\vec{n},\vec{n}^1\}$.  Each depicted structure is just one of
    the many representatives of its class, as continuous rotations
    preserve the topology of the defects.  (e,f) Numerical model of
    blue phases I and II with isosurfaces representing half-integer
    native (red) and quasi-(cyan) disclinations. The blue phase I
    has the quasi-disclinations running along the axes of double
    twist cylinders, extending infinitely along each lattice
    direction. The blue phase II has four native disclinations and
    four quasi-disclinations meeting at the same point, spanning
    all the diagonals of the cube in an alternating order. (g) At the
    quasi-disclinations, the axes $\vec{n}^1$ (red) and
    $\vec{n}^2$ (green) rotate by $-\pi$ around $\vec{n}$ (blue). (h)
    At the native disclinations, the $\vec{n}$ and $\vec{n}^1$ axes
    show a similar behavior. (f,g,h) In the blue phase II, exchanging
    the director $\vec{n}$ with $\vec{n}^2$ converts one disclination
    into the other and effectively rotates the unit cell by 90
    degrees. 
  }
\end{figure*}

First, we study the native nematic disclinations. They all involve the
director $\vec{n}$ as one of the degenerate axes. Moreover, on a tight
circle around the disclination, the director rotates very rapidly in
the plane defined by the two degenerate axes, so the singular value
associated with this direction {\sl diverges} at the defect
core. Since the singular values are sorted by magnitude, all native
nematic disclinations are, by construction, degenerate in the axes
$\{\vec{n}$, $\vec{n}^1\}$.  To see this, consider a plane
perpendicular to the defect line.  In this plane $\vec{n}$ winds
around ever more rapidly as we approach the core.  By definition, this
rapid winding is into the $\vec{n}^1$ direction at all points,
$w^1_i\partial_i n_j = \sigma_1 n_j^1$ with $\sigma_1$ diverging. Note
that since $\vec{n}\cdot\vec{n}^1=0$, $w^1_i\partial_i n^1_j = -
\sigma_1 n_j + \gamma n^2_j$ where $\gamma$ is finite.  It follows
that the winding is between $\vec{n}$ and $\vec{n}^1$ and that the
defects associated with the degeneracy of the smallest two singular
values, $\sigma_2=0$, must be topologically trivial in the $\vec{n}$
frame. The disclinations with a half-integer winding number, already
known from the homotopy theory of the bare director field, now also
have the perpendicular axis $\vec{n}^1$ performing a half-integer turn
(Fig.~\ref{fig:fig_blues}a). The addition of the perpendicular axes
also reveals disclinations with an odd integer winding number, which
are not topologically distinguished in the standard uniaxial setting,
but are well-defined in our definition (Fig.~\ref{fig:fig_blues}b).

The line-like quasi-disclinations, on the other hand, have smooth
director complexions, and so the degenerate axes with nonzero winding
number must be the invisible perpendicular axes
$\{\vec{n}^1,\vec{n}^2\}$. These disclinations are located where the
singular values $\sigma_1=\sigma_2$, precisely when we can no longer
distinguish the two axes in the decomposition (\ref{eq:svd}).  These
defects can have either half-integer or integer winding number
(Fig.~\ref{fig:fig_blues}c,d).

The odd integer-winding-number disclinations all belong to the same
$-\bone$ class of biaxial disclinations and can transform one into the
other: an unescaped integer winding number disclination with a
singular core in $\vec{n}$ can escape into the third dimension, which
removes the singularity, but still shows up as a defect in the
perpendicular axes $\{\vec{n}^1,\vec{n}^2\}$ of the frame. Our
construction therefore unambiguously locates both the escaped and
unescaped integer nematic disclinations, which would be impossible to
locate from the director field alone.

What about point defects in the uniaxial phase? Biaxial nematics
cannot have point defects as $\pi_2[{\rm SO}(3)/D_{2}]=0$.  As we
mentioned, to consider the biaxial structure we will inflate all point
defects into small disclination loops carrying the nontrivial element
of $\pi_1(\mathbb{R}P^2)$.  Each native disclination loop can either
carry an odd or even topological point charge, measured by the second
homotopy group, and it can be linked by an even or an odd number of
other disclination loops
\cite{trebin_rev,janich,nakanishi,randy_rmp}. The half-integer native
disclinations and all flavors of integer disclinations form an abelian
subgroup of the quaternion group $\jj^\nu$ with
$\nu\in\mathbb{Z}_4$. This periodicity of four, consistent with the
theoretical result given by the torus homotopy group
\cite{janich,randy_rmp} is seen in the specialized form for $-1/2$
disclination loops and their self-linking numbers
\cite{copar_rewiring}, and in the generalization to disclination loops
with arbitrary cross section \cite{copar_quaternion}. The orientation
of the disclination profile in the immediate neighborhood of the
singularity holds the information about the two degenerate axes that
vary rapidly around the singularity, foreshadowing the significance of
quaternions and differential definition of our biaxial framing. The
statement that the nematic hedgehog charge is a residual of linking a
$-\bone$ biaxial disclination \cite{randy_rmp} is intrinsic in our
construction.

Unlike in proper biaxial nematics, the point defects are still present
in our system -- as small native disclination loops, linked by
quasi-disclinations. To further explore the connection between the
point charge and the threading of the native loops, recall that the
topological charge $q$ in nematics is the degree of mapping
$\mathbb{R}^3\rightarrow \mathbb{R}P^2$ from a closed measuring
surface in the nematic medium to the unit sphere. The director
$\vec{n}$ on this enclosing surface is in fact a $q$-covered sphere,
which by the Riemann-Hurwitz theorem has the Euler characteristic of
$\chi=2q$. The transverse axes $\{\vec{n}^1,\vec{n}^2\}$ form a
tangent plane to $\mathbb{R}P^2$ and so, by the Poincar\'e-Brouwer
theorem, the total winding number of their surface defects must be
equal to the Euler characteristic. Every sphere that encloses a point
defect is penetrated by quasi-disclinations, with the sum of their
winding numbers equal to $2q$. All quasi-disclinations terminate on
the point defects and thread native disclination loops that carry a
topological charge.

Similarly to the native disclinations, the half-integer
quasi-disclinations form a distinct $\kk^\nu$ subgroup, with the same
algebraic structure and linking rules that we found for the native
nematic disclinations. However, the coexistence of half-integer
disclinations of both flavors reveals the nonabelian nature of the
quaternion group: the $\jj$ and $\kk$ disclination loops cannot link
without creating a $-\bone$ line connecting the loops
\cite{kleman_tether}. An additional feature of our gradient framing is
the {\sl absence} of $\ii$ quaternions representing the impossible
disclinations with a degenerate pair of axes $\{\vec{n}^2,\vec{n}\}$,
as discussed above.  This is a variation of Po\'enaru's theorem
\cite{poenaru} that limits the merging of defects in gradient fields,
as in smectics \cite{PNAS_cak}. Were a half-integer quasi-disclination
($\kk$) to merge with a half-integer native disclination ($\jj$), the
resulting defect would have the impossible signature $\ii$, so the
quasi-and native defects naturally avoid each other. They can only
meet at discrete points -- at point defects, which we have shown act
as sources or sinks for the quasi-disclinations.  As in systems with
broken translational symmetry \cite{mermin,PNAS_cak}, the missing
quaternion causes the homotopic description to be incomplete in this
case: the $\pm \ii$ elements of the fundamental group have no
realization in the sample.

Blue phases are a suitable system for studying the biaxial defects on
practical data, as they show nonuniform behavior without complicated
boundary conditions. We use the finite difference relaxation method,
based on the Landau-de~Gennes model used in
Ref.~\cite{pnas_ravnik}. The Q-tensor field was used to retrieve the
order parameter $S$, and the director field $\vec{n}$, which was
subsequently differentiated and decomposed with SVD, giving singular
values and framing information for each point in space. A large
resolution of $80$ points along each direction was used, as the
positions of the singularities are sensitive to errors caused by
finite difference approximation of the derivatives. The zeroes of the
order parameter $S$ were used to determine the position of the native
disclinations, while the zeroes of a functional $\sigma_1-\sigma_2$
were used to find the quasi-disclinations.

Both blue phases form a periodic cubic lattice. The blue phase I
consists of straight native disclinations, extending infinitely in the
direction of body diagonals and offset by half of the cell spacing to
avoid each other. The rest of the bulk can be roughly explained as
three mutually perpendicular double-twist cylinders, extending in the
direction of main coordinate axes \cite{wrightmermin}. Plotting the
near-zero isosurfaces of $\sigma_1-\sigma_2$ reveals three mutually
perpendicular infinitely extending quasi-disclinations that
approximately follow the double-twist cylinders, which is not
unexpected, as $\sigma_1=\sigma_2$ condition implies the rate of
change is equal in two directions, which is likely to occur near the
axis of a double twist cylinder (Fig.~\ref{fig:fig_blues}e).  In
particular, the geometric pattern of the cross section has a
half-integer winding number and looks like a characteristic three-fold
profile of a $-1/2$ nematic disclination line in the axes $\vec{n}^1$
and $\vec{n}^2$.

The blue phase II consists of a cross-linked network of
disclinations. The unit cell contains two junctions where four
diagonal native disclination lines meet in a tetrahedral
formation. The quasi-disclinations also extend from one junction to
the other in straight diagonal lines, forming a tetrahedral structure,
dual to the native one (Fig.~\ref{fig:fig_blues}f). The junctions are
thus highly degenerate, with four native and four half-integer
quasi-disclinations meeting there and extending to all 8 vertices of
the unit cube. An inspection of the disclination cross sections again
reveals that both the native and the quasi-disclinations have a
three-fold profile (Fig.~\ref{fig:fig_blues}g,h). Furthermore, the
cross sections of both disclination types are exact copies of each
other, with the axes $\vec{n}$ and $\vec{n}^2$ in exchanged roles. In
fact, the framing across the entire space possesses such a symmetry,
that an exchange of these axes has the same effect as a rotation of
the unit cell by $90^\circ$, which implies that even though the point
symmetry group around the central point is the tetrahedral group,
there are hints of the full cubic symmetry in the system. The
$\vec{n}^1$ axis in fact has a full cubic symmetry, up to a small
perturbation that depends on the difference in the free energy costs
of native and quasi-defects and could therefore be used as a model
director field for a periodic cubic structure with an eight-way
central junction of half-integer disclinations \cite{dupuis}. In
general, every director field has adjoint fields $\vec{n}^1$ and
$\vec{n}^2$ with equal or higher symmetry as the original director,
which have a potential use as model director fields for related
problems, as they already respect the boundary conditions and
approximate energy constraints.

In this paper, we explored the uniaxial nematic as a special kind of
biaxial nematic with hidden perpendicular degrees of freedom. Instead
of using the eigenvectors of the tensorial order parameter, which is
highly degenerate in the uniaxial phase, we retrieved the missing
perpendicular axes by using the SVD decomposition to extract a
smoothly varying frame from the director derivative, treating the
disclination lines as simple defects in the biaxial frame. This frame
is nevertheless related to the frame, retrieved from the $Q$-tensor,
as the first encodes the spatial variations of the director and the
latter describes thermal fluctuations of the molecular director.
Continuity of the director field and integrability conditions near
singular defects give rise to rules that restrict the set of allowed
defects, resulting in an intricate structure that is similar, yet not
equivalent to that of general biaxial defects. Beside the native
uniaxial nematic disclinations, known from the conventional homotopy
analysis, we uncover another type of disclinations that arise through
hidden biaxial order -- topologically unavoidable patterns in the
elastic response, such as the escaped integer disclination lines,
which were impossible to treat topologically in bare nematic director
fields, but can now be computationally detected. With the elusive
defects pinpointed deterministically, the complete set of linking
rules for disclination lines emerges and unifies the line and point
defects under the same formalism.

The SVD decomposition has a potential use in numerical and analytic
calculations, as it connects the free energy with topology via the
singular values. The $\vec{n}$-frame we focused on in this paper, can
be taken as a convenient choice of frame in the Mermin-Ho construction
\cite{kamien_primer}. Additionally, the conjugate $\vec{w}$-frame also
constitutes a biaxial frame that can be investigated in the future. As
a visualization technique, the singular values help to locate the
escaped defects and other characteristic features without resorting to
visualization of vector fields.

The technique illuminates an intricate link between geometry and
topology independently of the physical meaning of the order
parameter. We can find an underlying biaxial field and extract hidden
degrees of freedom for a wide variety of materials that allow
parametrization by an orthogonal frame. Of particular interest for
future research are patterns and defects in chiral nematics, smectic
liquid crystals and fields of polarized light, building on the
connection between local differential structure and the traditional
homotopy of defects.

The authors acknowledge support from Slovenian Research Agency under
Contracts No. P1-0099, No. J1-2335, NAMASTE Center of Excellence,
HIERARCHY FP7 network 215851-2 (S.\v{C}. and S.\v{Z}.), NSF Grant
DMR05-47230 (S.\v{C}. and R.D.K.) and Royal Society University
Research Fellowship (M.R.D.). This work was partially supported by NSF
Grant PHY11-25915 under the 2012 KITP miniprogram ``Knotted
Fields''. Special gratitude goes to T.~Porenta and M.~Ravnik, who
provided numerical data for the models of the blue phases.

\end{document}